\documentclass [12pt,twoside]{article}           % version LATEX2E

\usepackage{epsfig,times,lscape}
\usepackage[usenames]{color}

    \ifnum\count102=1

    \fi

    \ifnum\count102=2
\fi

\newcommand{\nc}{\newcommand}

     \ifnum\count101=1
\nc{\qI}[1]{\section{{#1}}}
\nc{\qA}[1]{\subsection{{#1}}}
\nc{\qun}[1]{\subsubsection{{#1}}}
\nc{\qa}[1]{\paragraph{{#1}}}

\def\qpar{\vskip 2mm plus 0.2mm minus 0.2mm}
\def\qL{\hfill \break}
     \fi 

      \ifnum\count101=2
 \nc{\qI}[1]{\parindent=0mm \vskip 8mm 
{\centerline{\LARGE \color{red}#1}}\vskip 3mm}
\nc{\qA}[1]{\vskip 2.5mm \noindent 
{{\bf\large\color{blue}  #1}} \vskip 1mm \parindent=0mm}
 \nc{\qun}[1]{\vskip 1mm \noindent {\sl #1 }\quad }

\def\qL{\hfill \break}
\def\qpar{\vskip 2mm plus 0.2mm minus 0.2mm}

      \fi
\def\qth{\vrule height 12pt depth 0pt width 0pt}
\def\qtb{\vrule height 0pt depth 5pt width 0pt}

\nc{\qfoot}[1]{\footnote{{#1}}}

      \ifnum\count101=1
\def\qbu{\hfill \par \hskip 6mm $ \bullet $ \hskip 2mm}

      \fi
      \ifnum\count101=2
\def\qbu{\hfill \par \hskip 4mm $ \bullet $ \hskip 2mm}

      \fi

\def\qparr{ \vskip 1.0mm plus 0.2mm minus 0.2mm \hangindent=10mm
\hangafter=1}

     \ifnum\count101=1 
 
     \fi
     \ifnum\count101=2

  \def\qcitb#1{\noindent \hbox to 102mm{\hfill \small #1} \vskip 1mm}
      \fi

 \def\qpages#1{\count102=0{\loop\advance\count102 by 1
 \null \vfill\eject \ifnum\count102<#1 \repeat}}

\def\qn#1{\eqno \hbox{(#1)}}

\def\qth{\vrule height 12pt depth 0pt width 0pt}
\def\qtb{\vrule height 0pt depth 5pt width 0pt}

\def\qv{\vskip 0.1mm plus 0.05mm minus 0.05mm}
\def\qhu{\hskip 0.6mm}
\def\qhv{\hskip 3mm}

\def\qhw{\hskip 1.5mm}
\def\qleg#1#2#3{\noindent {\bf \small #1\qhw}{\small #2\qhw}{\it \small #3}\qv }

\begin{document}
\thispagestyle{empty}
% --------------------------------------------------------------------

      % Hauts de pages et numerotation

          % Remarque: sans le \protect --> message d'erreur (ordre fragile)
\markboth{{\sl \hfill  \hfill \protect\phantom{3}}}
        {{\protect\phantom{3}\sl \hfill  \hfill}}

% -------------------------------------------------------------------
\color{yellow} 
\hrule height 20mm depth 10mm width 170mm 
\color{black}
\vskip -2.5cm 
\centerline{\bf \Large Incidence of the Bertillon and 
Gompertz effects on}
\vskip 2mm
\centerline{\bf \Large the outcome of clinical trials}
\vskip 4mm

\centerline{\large 
Bertrand M. Roehner$ ^{1} $
}

\vskip 15mm
\large

{\bf Abstract}\quad
The accounts of medical trials provide very detailed
information about the patients' health conditions.
On the contrary, 
only minimal data are usually given about demographic factors. 
Yet, some of these factors
can have a notable impact on the overall death rate,
thereby changing the outcome and conclusions of the trial.
This paper focuses on two of these variables.
The first is marital
status; this effect, which
will be referred to as the Bertillon effect, 
may change death rates by over 100\%
The second is the age of the oldest patients;
because of the exponential nature of Gompertz's law, 
the distribution of ages in the oldest age group can
have dramatic consequences on the overall number of deaths.
It will be seen that randomization alone can hardly take care
of these problems. Appropriate remedies are easy to formulate 
however. 
First, the marital status of patients as well as the 
age distribution of those over 65 should be documented
for both study groups.
Then, thanks to these data and based on the Bertillon 
and Gompertz laws,
it will become possible to perform appropriate
corrections.
Such corrections will notably improve the reliability
and accuracy of the conclusions, especially in trials which
include a large proportion of elderly subjects. 

\vskip 6mm
%\centerline{\it First version: 9 June 2013, comments are welcome}
\centerline{\it First version: 9 June 2013}

\vskip 15mm
{\normalsize Key-words: Gompertz law,
median age, randomized subjects
mortality, marital status,
clinical trials, bias.}
\vskip 15mm

%PACS classification: Interdisciplinary (89.20.-a) +
%collective effects (71.45.Gm)
%\vskip 8mm

{\normalsize 
1: Institute for Theoretical and High Energy Physics (LPTHE),
University Pierre and Marie Curie, Paris, France. \qL
Email: roehner@lpthe.jussieu.fr
}

\vfill\eject

\qI{Overview}

In recent decades clinical trials have become
highly technical and standardized procedures
There is even a scale, the Jadad scale, which 
 assesses the methodological quality of a clinical trial,
particularly regarding randomization and blinding.
The accounts commonly contain the following sentence:
``The participants in the two study groups
[i.e. placebo versus drug group] were well balanced with respect to
major risk factors''. In support of this claim the papers 
provide a table entitled ``Baseline characteristics of the
trial participants according to study group''.
Table 1 reproduces the non-medical factors as given
in the account of the LIPID (1998) trial. Other accounts (e,g.
Jupiter 2008, WOSCOPS 2007) contain similar tables.
\qpar

Yet, it seems that two important factors are commonly omitted
which can
substantially affect the outcome in terms
of overall death rate%
\qfoot{In this paper we focus on clinical trials in which the
number of deaths occurring in each group (placebo
versus drug) is a key-result.
This is for instance the case of trials involving
drugs for the treatment of heart diseases.}%
,
namely the marital situation ($ MS $) of the patients 
and the age distribution of the fraction older than
65 ($ F65 $). 
\qbu The first point is related to the fact that the
death rate by heart disease or by cancer is highly dependent
(by a factor 2 as shown in Fig. 1)
upon the marital status of the subjects.
For the sake
of brevity this effect will be referred to as the Bertillon effect%
\qfoot{After Louis-Adolphe Bertillon (1872) who stated
the rule for overall death-rates in every age interval above 20.
Subsequently he reported
(Bertillon 1879, p.474) a  similar observation for  
suicide rates.In this case the effect
is about 1.5 times stronger. Some twenty years later
his study was revisited and expanded by Emile 
Durkheim (1897, Part 2, Chapter 3).}%
.
\qbu Most accounts of clinical trials give the median age 
($ MA $) 
in the placebo ($ P $) and drug ($ D $) groups.
It will be shown that the median age is a very poor
indicator of overall expected mortality in any group of people.
This is due to 
Gompertz's law for death rates according to which
death rates increase as an exponential function of
age%
\qfoot{For ages over 35 and in the conditions of medical trials
one can neglect the age-independent Makeham component
of the mortality rate.}%
.
Roughly speaking,
after the age of 35 the age-specific death rate doubles
every 10 years. In the age group 80-84 it is 40 times higher
than in the age group 35-39. Therefore it is not surprising
that the fraction of elderly people is a much better
indicator of expected death rate (see Fig. 4).
\qpar

One could argue that the previous observations do not
really matter because the randomization procedure will take
care of that and ensure that the $ P $ and $ D $ groups are
identical with respect to $ MS $ and $ F65 $.
This is not true however. 
\qpar

Randomization may indeed 
result in groups having  similar $ MA $  because 
in any set of groups this variable has a low dispersion.
However, this does not tell us anything significant about
death rates. In contrast, the more significant
$ F65 $ variable has a much higher dispersion.
For a given sample of US counties
its coefficient of variation ($ CV=
\sigma/m $, is 3 times the $ CV $ of $ MA $.
For $ MS $ the randomization is also a tricky operation
because there are 5 different marital situations,
namely: single, married, non-married partners, 
divorced, widowed, each of 
which has a different expected mortality.
As the rates are different for males and females one needs
in fact 10 categories. 
Thus, it will be nearly impossible to have balanced
population numbers for each of these groups.
This problem becomes more serious as the average age
of the study group increases. At age 50, 85\% of the US population
was married (in 1980), but for the age group 
over 75 years, 46\% were married, 45\% 
widowed, 5\% single and 2\% divorced (Statistical Abstract of the
US 1981, Table 49).
\qpar

\qI{Incidence of the Bertillon effect}

Although, as already mentioned, the influence of marital status
on death rates has been known for a long time,
knowledge of this effect separately for different
causes of death (as summarized in Fig. 1) is more recent. As a
matter of fact, such evidence can only be obtained in a country
such as the United States which has a large population.
For instance, in 1979 the number of widowed males 
who died from cancer in the age group 35-44 was only 62 for the 
whole country; in 1980, it was 48 which shows that despite
being small these death numbers are nevertheless fairly stable
(Vital Statistics of the US, 1979 and 1980).
 
%
%%%% MORTALITE ``HEART+CANCER'' PAR ``MARITAL STATUS''
\begin{figure}[htb]
\centerline{\psfig{width=14cm,figure=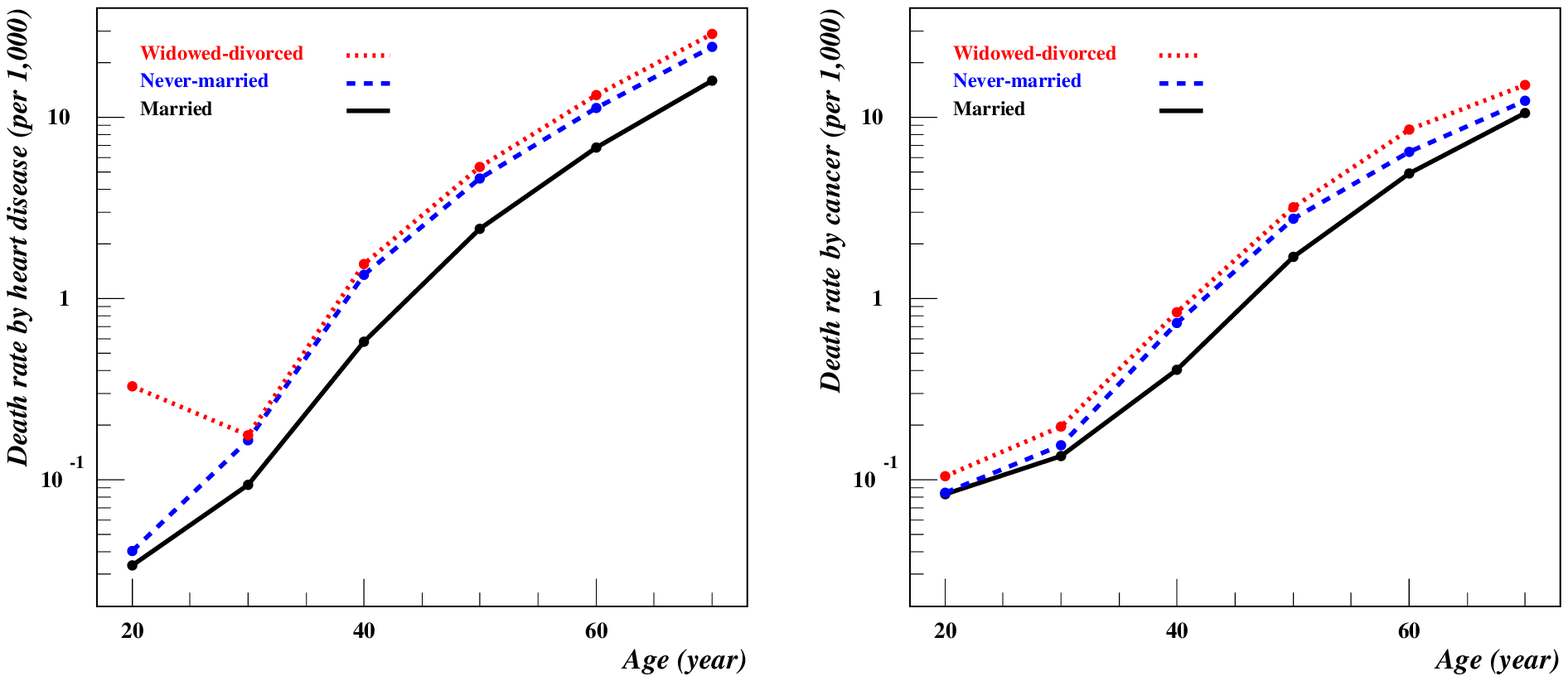}}
\qleg{Fig.\qhu 1 a,b\qhv Death rate through heart disease or cancer
by marital status, males, United States, 1980.}
{The death rate of married people is on average two times smaller
than the death rate in the two other categories.
More precisely, the  ratios single/married and
widowed/married are as follows: \qL 
$ \hbox{age 40, heart: }  s/m=2.3,\ w/m=2.7;\ 
\hbox{cancer: } s/m=1.8\ w/m=2.1 $ \qL
$ \hbox{age 50, heart: }  s/m=1.9,\ w/m=2.2;\ 
\hbox{cancer: } s/m=1.6\ w/m=1.9 $ \qL
For the sake of graphical clarity,
widowhood and divorce cases were lumped together.
However, a more detailed analysis
shows that widowed people have 
markedly higher death rates than divorced persons.}
{Sources: Number of deaths: Vital Statistics of the United States, 1980,
Vol.2: Mortality, Part A, p.316-317. Population by age and marital
status: 1980 Census Census of population, Marital Characteristics,
p.1; Statistical Abstract of the United States 1981, Table 49.}

\end{figure}
%----------------------------------------------

As noted above, randomization will hardly be able to
balance exactly the proportions of different MS in the study groups
but it may also be argued that randomization is not likely
to result in {\it huge} imbalances. 
However, such imbalances may be produced indirectly.
For instance, if one wishes to compare groups of males and females
over 65 years
one must be aware that the proportions of widowed subjects will
be very different in the two cases: 
about 8\% for males compared to 40\% for
females (Statistical Abstract of the United States 1980).
\qpar

As the 
order of magnitude of the Bertillon effect is around 100\%, even
a partial imbalance can matter 
because the difference in number of deaths between  
$ P $ and $ D $ groups
is usually of the order of 20\%-30\%.

\qI{Incidence of the Gompertz effect}

This section proceeds in three steps. Firstly, the Gompertz law is
recalled. Secondly, We present a thought experiment which
conveys the main idea. Thirdly, we emphasize that the population
fraction over 65 is a much better predictor of the overall death
rate than the median age.

\qA{Gompertz law}

Discovered in 1824, Gompertz's law was probably the first
major law in the field of demography. As shown in Fig. 2, it is not
only valid for overall death rates but also separately
(with only slight variations) for different diseases. 

%
%%%% LOI DE GOMPERTZ
\begin{figure}[htb]
\centerline{\psfig{width=11cm,figure=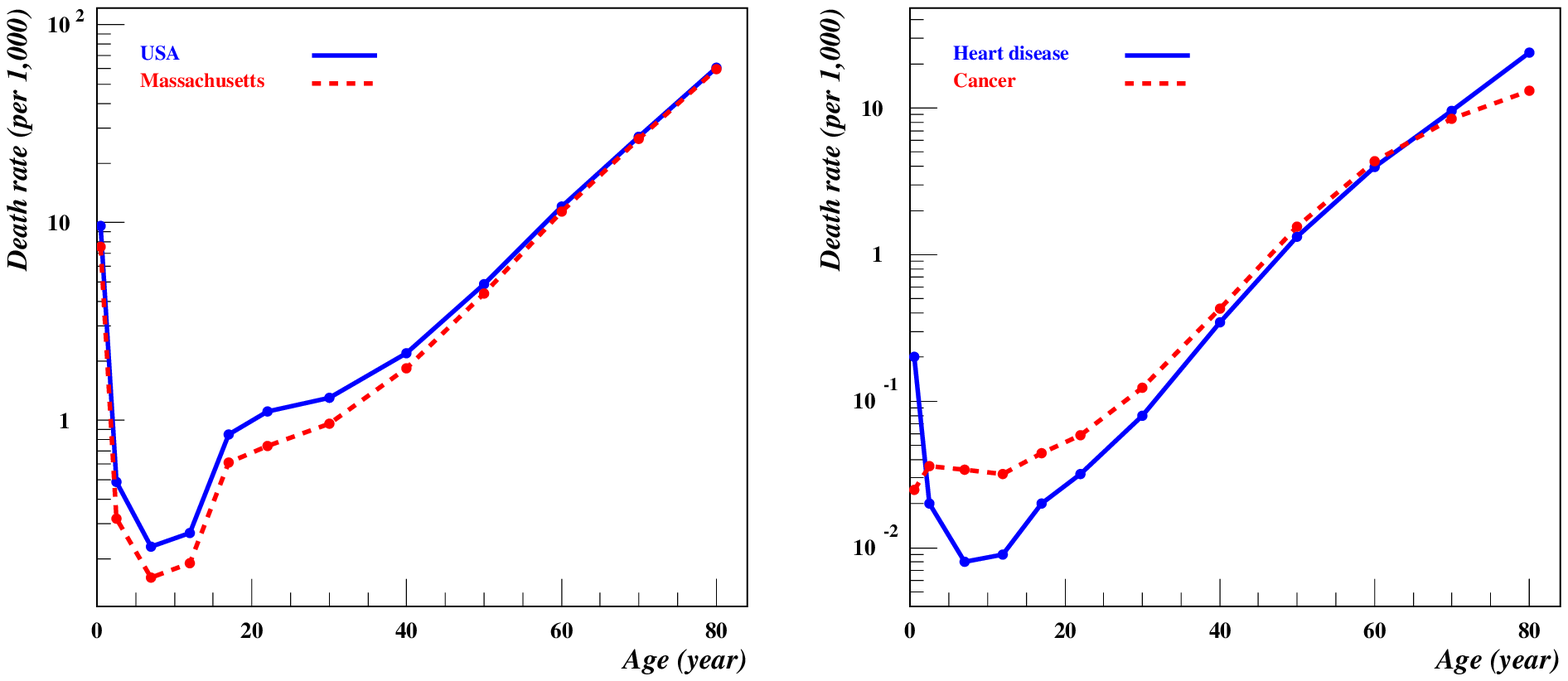}}
\qleg{Fig.\qhu 2a,b \qhv Age-specific death rates in the United States
in 2000.}
{The graph illustrates the exponential growth of death rates
as a function of age which constitutes Gompertz's law.
Not surprisingly,
there is a similar exponential growth for {\it separate} causes
of death provided they are due to diseases 
(e.g. cancer, heart disease, cerebrovascular accident, pulmonary
disease, etc.).
On the contrary, the rates for causes of death 
that are not diseases (e.g. suicide, accident)
do {\it not} increase exponentially.}
{Source: Centers for Disease Control and Prevention, National
Center for Health Statistics, Compressed Mortality File. 
This database is commonly referred to as the WONDER database.}
\end{figure}
%----------------------------------------------

In what follows we will adopt the following parameters:
 $$ y=g_0\exp(ax)\quad
x:\hbox{ \small age (in years)},\ 
y:\hbox{ \small death rate per 1,000},\ a=0.082,\ g_0=0.11 \qn{1} $$

It will be seen below that they lead to death number predictions 
which are consistent with what is observed in medical trials.

\qA{Thought experiment based on LIPID (1998)}
The LIPID (1998) experiment was selected because it provides
much more information than usually given about the
demographic characteristics of the study groups.
Whereas many accounts give only the median age,
 LIPID (1998) gives 6 characteristics
which are recalled in Table 1 and Fig. 3. Yet,
on the basis of Gompertz's law one quickly comes to
realize that these characteristics are not sufficient
and not well chosen. They are not sufficient
because they do not give enough information about
the oldest subjects in each group. For instance, 
the account says that the subjects' age interval was 31 to 75
but it does not give the intervals separately for the two
study groups. Moreover, the characteristics that are
given do not put any constraint on the distribution of
ages in the {\it oldest} age group.

%%-----------------------------------------------
\begin{table}[htb]

\centerline{\bf  Table 1\ Base-line characteristics of patients
in the LIPID (1998) trial.}

\vskip 5mm
\hrule
\vskip 0.5mm
\hrule
\vskip 2mm

\color{black} 

$$ \matrix{
\qtb
\hbox{Age characteristic} \hfill  &
\hbox{Placebo group} & \hbox{\phantom{aa}}& \hbox{Pravastin group}\cr
\noalign{\hrule}
\qth
\hbox{All ages} \hfill & 4,502 \hfill 100\% & &4,512 \hfill 100\% \cr
31-54 \hfill &       1,021 \hfill 23\% & &1,065 \hfill 24\% \cr
55-64 \hfill & 1,708 \hfill 38\% & &1,706 \hfill 38\% \cr
65-69 \hfill & 1,087 \hfill 24\% & &1,081 \hfill 24\% \cr
70-75 \hfill & 686 \hfill 15\% & & 660 \hfill 15\% \cr
 & & &\cr
\hbox{Median age}\hfill & 62 & &62 \cr
\hbox{Interquartile range} \hfill & 55-68 & & 55-67 \cr
 & & & \cr
\qtb
\hbox{Deaths from any cause (6 years)} & 633 & & 498  \cr
\noalign{\hrule}
} $$
\vskip 0.5mm
Notes: The table is reproduced by keeping all figures
in the very same form as given in the source.
Two observations are in order. (i) Although 
several statistical
characteristics are given, they all fail to describe
the 4th quartile,
i.e. the fraction of the 25\% oldest people in each group
(which comprises the age-group 70-75 because 686/4502=15\%)
The thought experiment delineated in Fig. 3
shows that these characteristics
greatly matter as far as the
overall expected mortality is concerned.
(ii) In the source
the fraction 70-75 was rounded to the closest integer, namely 15\%.
More precise values are $ F70=15.24\%,\ 14.63\% $.
The difference 686-660 will result in an expected number
of 6 more deaths in the placebo group which represents 4.4\% of the
overall death difference of 135 between the two
groups. Although small, this correction should not
be omitted.
\qL
Source: LIPID (1998).
\vskip 2mm
\hrule
\vskip 0.7mm
\hrule
\end{table}
%%-----------------------------------------------

%
%%%% ``THOUGHT EXPERIMENT'' A PARTIR DE L'ESSAI LIPID (1998)
\begin{figure}[htb]
\centerline{\psfig{width=9cm,figure=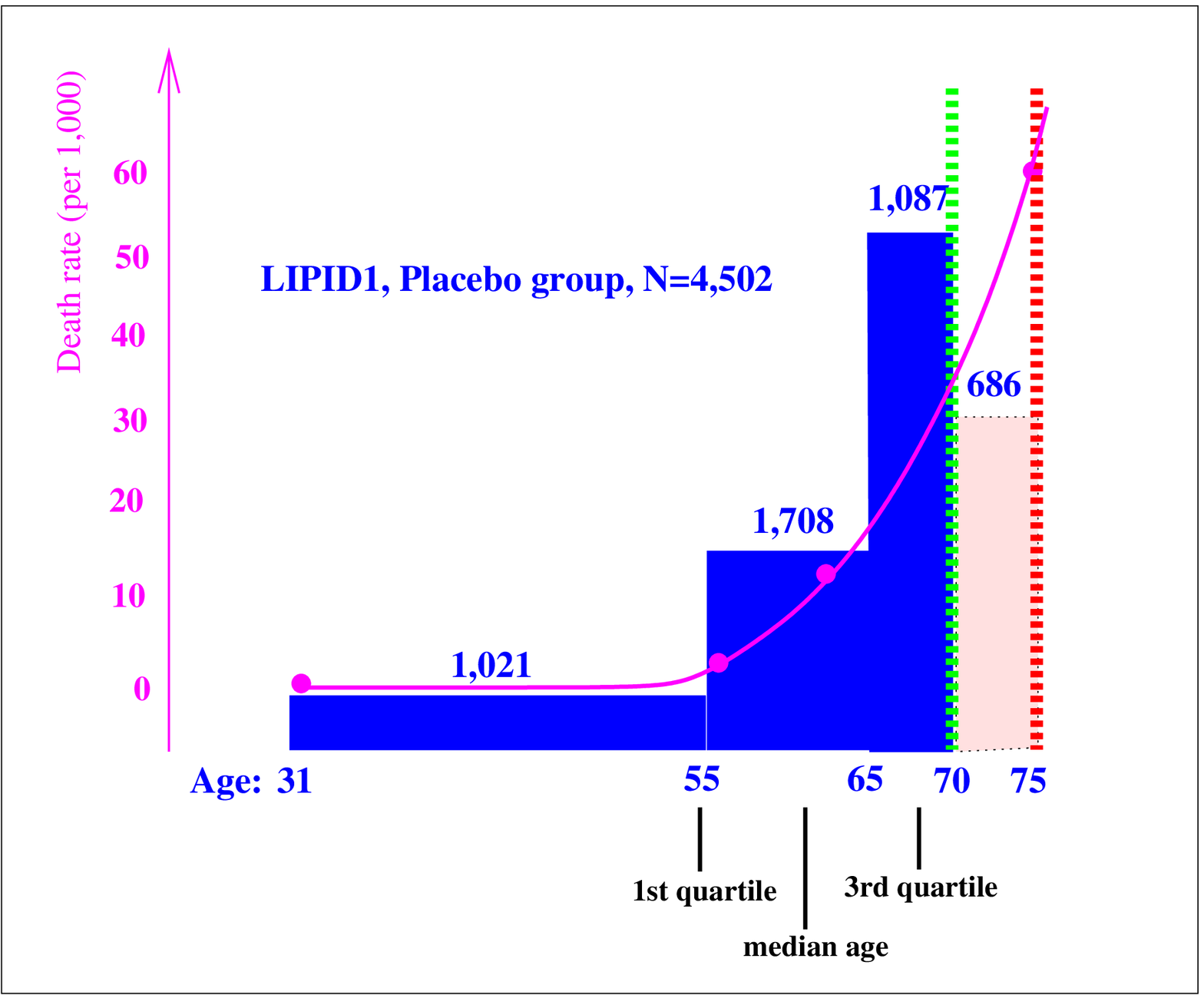}}
\qleg{Fig.\qhu 3\qhv Changes that do not
affect the reported statistical characteristics
but significantly change the overall number of deaths.}
{The histogram 
corresponds to the age groups of the LIPID (1998) trial. 
The oldest age group has been drawn in a different color
because the statistical data reported in the paper
(and reproduced in Table 1) do not put any constraint on the
distribution of subjects {\it within} this group.
Three cases are represented: uniform distribution (pink),
only 70-year old subjects, only 75-year old subjects.
Depending on the assumption, the
number of deaths (all causes) for the whole trial 
may change by as much as 11\%.
In LIPID (1998) the oldest age group represented 
only 15\% of the whole study groups;
needless to say, the Gompertz effect will be stronger when
this percentage is higher. The effect becomes also stronger as
the age of the oldest age group increases.
Note that the scale on the left-hand side is for the death rate
curve; the scale for the histogram is not shown.} 
{Source: LIPID (1998)}
\end{figure}
%----------------------------------------------

Thus, by modifying this distribution one can  
substantially change the expected
number of deaths. 
\qpar
How did we carry out this analysis?
For each age equation (1) gives the corresponding death rate.
As the trial lasted 6 years, the calculation will involve 
the following steps.
If (e.g. in the $ P $ group) 
there are initially $ n_1 $ subjects aged 61, some $ m_1=n_1y(61) $
will die in the first year. As a result, at the beginning of year 2 
there will be $ n_2=n_1-m_1 $ remaining patients. 
Similarly, during the second year, some $ m_2=n_2y(62) $ will die.
By repeating this calculation first for the 6 years of the trial,
and then for all ages, and by summing all death numbers
one gets the expected death number during the whole trial.
In this way, one obtains the following results (this is
for the placebo group%
\qfoot{Because the age end-points (31,75) are only given for the
whole sample, we had to assume that they are identical for the
placebo group.}%
):
$$ 31-54:31,\enskip 55-54:180,\enskip 65-69:200,\enskip 70-75:190, 
\enskip \hbox{\normalsize total: } 601\hbox{ \small (instead of } 633) $$

The fact that the actual number of deaths, namely 633,
is slightly higher than
the expected number may be due to the selection of the patients.
All of them had a history of myocardial infarction and also a fairly high
cholesterol level. On account of this, 
for subsequent calculations the coefficient $ g_0 $ will
be multiplied by the following renormalization factor $ 633/601=1.053 $. 
\qpar

Now, we are ready to carry out the experiment described in 
Fig. 3. The number of deaths in the age group 70-75
were calculated under the following assumptions:
\qbu Whole group at age 70
\qbu Uniform distribution 
\qbu Whole group at age 75
\qL
and lead to the following results.

$$ \hbox{Age 70: } 164\hbox{ \small deaths},\quad 
\hbox{Uniform: } 198\hbox{ \small deaths},\quad 
\hbox{Age 75: } 235\hbox{ \small deaths} $$

The difference between the cases ``70'' and ``75'' is 71 deaths 
which represent 
11.2\% of the total number of 633 deaths in the placebo group.
This is one half of the
difference between the placebo and drug groups,
namely $ (633-498)/633=21.3\% $.

\qA{Determinants of death rate}

As a proof of the fact that the median age
is not a useful variable, we show that it is a poor predictor
of overall death rates. In contrast, $ F65 $
is a very good predictor of death rates.
\qpar

The data shown in Figure 4 are for the 159 counties of Georgia.
For what reason was Georgia selected?
Altogether there are some three thousand counties in the United
States but the numbers of counties per state vary greatly.
Texas and Georgia  are among the states with the
largest numbers of counties which is why they were selected.
%
%%%% COMPARAISON AGE MEDIAN ET F65
\begin{figure}[htb]
\centerline{\psfig{width=12cm,figure=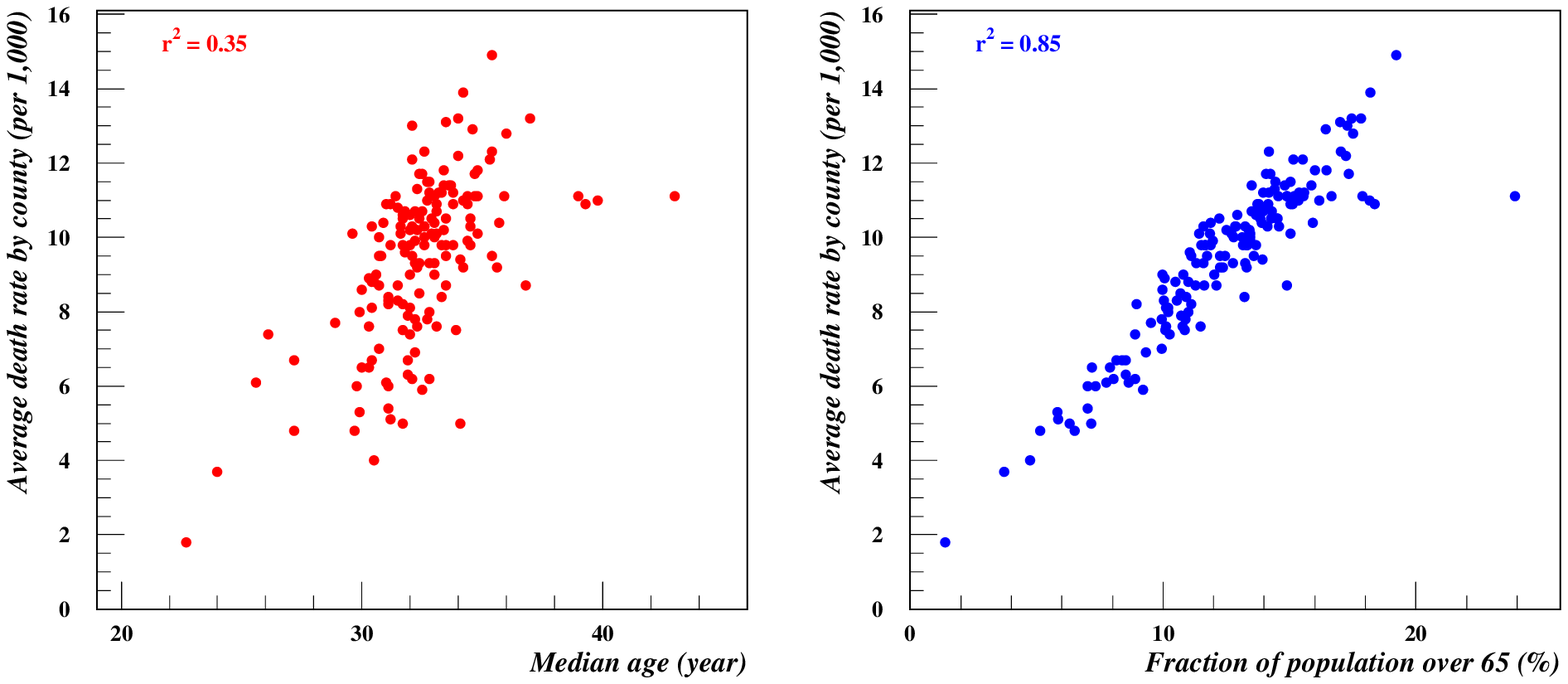}}
\qleg{Fig.\qhu 4 a,b\qhv Relationship between median age (left)
or fraction over 65 (right)
on the one hand and average death rates on the
other hand for the counties of Georgia.}
{The poor correlation between median ages and
death rates is a consequence of Gompertz's law. Indeed, the fact that
the death rate at age
82 is 40 times higher than at age 37 implies that the addition
of young persons will lower the mean age 
without notably changing the death rate. On the contrary, 
just a few more persons over 80 will lift the death
rate without substantially shifting the mean age.
The median age is even less sensitive to such changes
than the mean age.
Note that in the graph the median age 
and $ F65 $ are for the year 1990 whereas 
the death rates are averages over 1979-1998.
The least-square estimate of the regression line between $ F65 $
and the death rate $ d $ (per 1,000 population) reads:\qL
\centerline{$ d=aF65+b,\quad a=0.61\pm 0.04,\quad b=1.8\pm 0.13 $.}
There are similar results in other states; for instance
in Texas (254 counties) 
the $ r^2 $ of the correlation $ (F65,d) $
is 0.92 and the parameters
$ a,b $ of the regression line are: $ a=0.59\pm 0.02,\quad b=1.3\pm 0.11 $.}
{Sources: Average age: Bureau of the Census, USA Counties website.
F65 and average death rates:
Centers for Disease Control and Prevention, National
Center for Health Statistics, Compressed Mortality File (commonly
called ``WONDER'' database).}
\end{figure}
%----------------------------------------------

\qI{Conclusion}
In this paper
we emphasized that, due to the Bertillon effect, the marital status
of the subjects taking part in a trial is of cardinal importance
because it may increase death rates due to heart disease or cancer
by as much as 100\%. Although for the sake of brevity we focused
our attention on these two major causes of death, there is
a similar effect (of same magnitude) for other causes
of death such as cerebrovascular accidents or pulmonary diseases.
This observation leads to the recommendation to include
information about marital status in the table giving
the characteristics of the two study groups.
\qpar
Secondly, we emphasized that, due to the Gompertz effect,
it is important to describe the oldest fractions of the study
groups in great detail. In several study accounts we were not even able
to find the limits of the age intervals of the study groups.
The most appropriate information would be the density functions
of the age groups over 65 as a function of age. If, for some reason,
this is not possible then one should give at least the
percentage fraction over 65 ($ F65 $) and the age of the oldest
subjects in each group.
We have seen that the median age is almost useless because it is a poor
predictor of overall death rates. 
\qpar
By giving the possibility of performing appropriate
corrections, 
the two points made here should permit to improve the accuracy of
trial results. As this can be done at little cost in terms of
additional text in account papers there is really no reason
to discard such an improvement.

\vskip 5mm
{\bf References}

\qparr
Bertillon (L.-A.) 1872: Article ``Mariage'' in the
Dictionnaire Encyclop\'edique des Sciences M\'edicales,
2nd series, Vol. 5, p.7-52.

\qparr
Bertillon (L.-A.) 1879: Article ``France'' in the
Dictionnaire Encyclop\'edique des Sciences M\'edicales,
4th series, Vol. 5, p.403-584.

\qparr
Durkheim (E.) 1897: Le suicide. Etude de sociologie. F. Alcan, Paris
[A recent English translation is: ``On Suicide'' (2006),
Penguin Books, London.]

\qparr
JUPITER Study Group 2008: Rosuvastatin to prevent vascular events in men
and women with elevated C-reactive protein.
The New England Journal of Medicine 359,21,2195-2207.

\qparr
LIPID Study Group 1998: Prevention of cardiovacular events and death
with Pravastin in patients with coronary heart diseases.
The New England Journal of Medicine 339,19,1349.

\qparr
WOSCOPS Study Group 2007: Long-term follow-up of the West of Scotland
coronary prevention study.
The New England Journal of Medicine 357,15,1477-1486.

\end{document}